\documentclass[12pt]{article}
\usepackage{axodraw}
\usepackage{epsfig}
\def\beq{\begin{equation}}
\def\eeq{\end{equation}}
\def\bea{\begin{eqnarray}}
\def\eea{\end{eqnarray}}
\def\bq{\begin{quote}}
\def\eq{\end{quote}}

\def\bq{\begin{quote}}
\def\eq{\end{quote}}

\def\bq{\begin{quote}}
\def\eq{\end{quote}}

\def\gappeq{\mathrel{\rlap {\raise.5ex\hbox{$>$}}
{\lower.5ex\hbox{$\sim$}}}}

\def\lappeq{\mathrel{\rlap{\raise.5ex\hbox{$<$}}
{\lower.5ex\hbox{$\sim$}}}}

\def\bbz{fa Z \kern-8.9pt Z}

\begin{document}

\baselineskip 18pt
\newcommand{\sheptitle}
{Solving the BFKL Equation with Running Coupling}

\newcommand{\shepauthor}
{J.R. Forshaw$^{1}$, D.A. Ross${^2}$ and A. Sabio Vera${^3}$}

\newcommand{\shepaddress}
{${^1}$Department of Physics and Astronomy, University of Manchester,
 Oxford Road, Manchester M13 9PL, U.K.\\
${^2}$ Division Th\'{e}orique, CERN, CH1211-Geneva 23, Switzerland
\footnote{On Leave of absence from:\\Department of Physics
and Astronomy, University of Southampton, Southampton, SO17 1BJ,
U.K.} \\
$^{3}$ Cavendish Laboratory, University of Cambridge, Madingley Road, Cambridge, CB3 0HE, U.K.  }

\newcommand{\shepabstract}
{We describe a formalism for solving the BFKL equation with a coupling
that runs for momenta above a certain infrared cutoff.  By suitably
choosing 
matching conditions proper account is taken of the fact that the BFKL
diffusion implies that the solution in the infrared (fixed coupling) regime
depends upon the solution in the ultraviolet (running coupling) regime and 
vice versa.
Expanding the BFKL kernel to a given order in the ratio of the transverse 
momenta allows arbitrary accuracy to be achieved.}

\begin{titlepage}
\begin{flushright}
\verb=Cavendish-HEP-2000/11= \\
\verb=CERN-TH/2000-354= \\
\verb=MC-TH-00/11= \\
\verb=hep-ph/0011047=\\
\end{flushright}
\begin{center}
{\large{\bf \sheptitle}}
\bigskip \\ \shepauthor \\ \mbox{} \\ {\it \shepaddress} \\ \vspace{.5in}
{\bf Abstract} \bigskip \end{center} \setcounter{page}{0}
\shepabstract
\end{titlepage}

Diffractive processes are
described, at least within the framework of  perturbative QCD,
 by the BFKL equation \cite{bfkl}. This equation
sums the leading and next-to-leading $\ln s$ \cite{BFKL2} terms for the 
amplitude involving the exchange in the $t-$channel with the quantum numbers 
of the vacuum.

In the case of fixed coupling, the energy dependence of these diffractive
processes is determined by the emergence of a branch cut in the Mellin 
transform plane ($\omega-$plane), with a branch point at $\omega > 0$. 
Lipatov \cite{lipatov86} has demonstrated that 
accounting for the running of the strong coupling together with some
(non-perturbative) information about the infrared behaviour of QCD
leads to a separation of the $\omega-$plane singularity structure into a 
series of  isolated poles. 
More recently, Thorne \cite{thorne} has shown that if one 
uses the solution to the BFKL equation with running coupling, the factor in the
amplitude which is totally calculable in perturbative QCD has an essential
singularity at $\omega=0$, such that there is no power-like energy dependence.
This is in accord with the recent work of Ciafaloni, Colferai and Salam, who 
have shown that the leading $\omega$-plane singularity is of non-perturbative
origin \cite{ccs}.

The BFKL equation is a diffusion equation and so there will always be some
contribution from the infrared region. One way of treating this region
is to assume that the strong coupling ceases to run below a certain
scale. In this case, we expect that for sufficiently low values of the 
transverse momenta of the exchanged gluons, the $s-$behaviour will be 
dominated by  the fixed coupling solution, whereas for large enough transverse
momenta the solution of \cite{thorne} will dominate. We should like to 
understand the smooth extrapolation between these two extremes.

In this letter we describe a formalism for solving the BFKL equation
with strong coupling running above some infrared scale, in which we
can see how the transition from the fixed coupling to the running coupling
solution arises.
\bigskip

\noindent {\bf Formalism for Solving the BFKL Equation with Running Coupling}

To begin with we  consider  only the leading order
BFKL kernel and describe later how the formalism may be extended in order to
 include the substantial part of the higher order corrections.

Since the BFKL equation is a diffusion equation there is always
diffusion into the infrared regime, where renormalization group improved
perturbation theory breaks down and one needs supplementary information
about the behaviour of QCD beyond perturbation theory. For our purposes
we simulate this infrared behaviour by assuming that the coupling
freezes below a critical value of $t \equiv \ln(k^2/\Lambda_{QCD}^2),
\ t=t_0$. We also make the assumption
that $\alpha_s(t_0)$ is sufficiently small that perturbative results
in the infrared regime maintain some level of credibility.

The full BFKL equation is solved in terms of a complete set of eigenfunctions
$f_\omega(t)$ of the kernel with eigenvalue $\omega$. For any $\omega$
there will be some diffusion into the infrared regime. We therefore
write $f_\omega(t)$ as
\beq f_\omega(t) \ = \ f_\omega^<(t) \, \theta(t_0-t) +
   f_\omega^>(t) \, \theta(t-t_0) \eeq
and the kernel as
\begin{eqnarray}
K(t,t^\prime) &=& \left[ \frac{1}{b \, t_0} {\cal K}^<(t,t^\prime)
     \, \theta(t_0-t') +  \frac{1}{b \, t }{\cal K}^<(t,t^\prime)
     \, \theta(t'-t_0) \right] \, \theta(t-t^\prime)  \ + \nonumber \\ & &  \ 
   \left[ \frac{1}{b \, t_0} {\cal K}^>(t,t^\prime)
     \, \theta(t_0-t') +  \frac{1}{b \, t^\prime }{\cal K}^>(t,t^\prime)
     \, \theta(t'-t_0) \right] \, \theta(t^\prime-t) \end{eqnarray}
where $b=\beta_0 /12$, $\beta_0=11-2n_f/3$, being the first coefficient of
the $\beta-$function for the running of the QCD coupling.
We define $ {\cal K}^<(t,t^\prime)$ and $ {\cal K}^>(t,t^\prime)$
as the limits of sums, i.e.
$$  {\cal K}^<(t,t^\prime) \ = \ \lim_{N \to \infty}  {\cal K}^<_N(t,t^\prime)
$$
$$  {\cal K}^>(t,t^\prime) \ = \ \lim_{N \to \infty}  {\cal K}^>_N(t,t^\prime)
$$
where
\beq {\cal K}^<_N(t,t^\prime) \ = \ \sum_{r=0}^N \left\{
 e^{(r+1/2)(t^\prime-t)}-\frac{1}{r+1} \delta(t-t^\prime) \right\} \eeq 
\beq {\cal K}^>_N(t,t^\prime) \ = \ \sum_{r=0}^N \left\{
 e^{(r+1/2)(t-t^\prime)}-\frac{1}{r+1} \delta(t-t^\prime) \right\}. \eeq 
The functions $e^{\gamma \, t}$ are eigenfunctions of these kernels
with eigenvalues $\chi_N^<(\gamma)$ and $\chi_N^>(\gamma)$ respectively, i.e.
\begin{eqnarray}
\int_{-\infty}^{t} {\cal K}^<_N(t,t^\prime) e^{\gamma t^\prime} dt^\prime
&=& \chi_N^<(\gamma) e^{\gamma t}  \nonumber \\ 
\int_{t}^{\infty} {\cal K}^>_N(t,t^\prime) e^{\gamma t^\prime} dt^\prime
&=& \chi_N^>(\gamma) e^{\gamma t}
\end{eqnarray}
where
\beq \chi_N^<(\gamma) 
\ = \ \sum_{r=0}^N \left\{ \frac{1}{r+1/2+\gamma}-\frac{1}{r+1}
\right\} \eeq
\beq \chi_N^>(\gamma)
 \ = \ \sum_{r=0}^N \left\{ \frac{1}{r+1/2-\gamma}-\frac{1}{r+1}
\right\}. \eeq
Note that $\chi_{\infty}^<(\gamma) = \gamma_E - \psi(1/2+\gamma)$ and
$\chi_{\infty}^>(\gamma) = \gamma_E - \psi(1/2-\gamma)$, and we have the
leading order BFKL kernel.

In this paper we truncate the kernel after $N$ terms. Note that the 
case $N=0$ reduces to the collinear model of \cite{ccs}.
For this truncated kernel the spectrum of eigenvalues is
\beq -\sum_{r=0}^N \frac{2}{r+1}\ \leq b \, t_0 \, \omega
\ \leq \sum_{r=0}^N \frac{2}{(2r+1)(r+1)}. \label{range} \eeq
The infrared part of the eigenfunction 
$ f_\omega^<(t)$ obeys the integral equation
\begin{eqnarray}
b \, t_0 \, \omega \,  f_\omega^<(t) & = & \int_{-\infty}^t 
{\cal K}^<_N(t,t^\prime)  f_\omega^<(t^\prime) \, dt^\prime 
 \ + \ \int_t^{t_0} 
{\cal K}^>_N(t,t^\prime)  f_\omega^<(t^\prime) \, dt^\prime \nonumber \\ & &
+ \ \int_{t_0}^\infty  {\cal K}^>_N(t,t^\prime)
 \frac{t_0}{t^\prime}  f_\omega^>(t^\prime)dt^\prime
\label{bfkl<} \end{eqnarray}
whereas the ultraviolet
part of the solution $f_\omega^>(t)$,  obeys the integral equation
\begin{eqnarray}
b \, t \omega \,  f_\omega^>(t) & = & \int_{t_0}^t 
{\cal K}^<_N(t,t^\prime)  f_\omega^>(t^\prime) \, dt^\prime 
 \ + \ \int_t^{\infty} 
{\cal K}^>_N(t,t^\prime)  \frac{t}{t^\prime} f_\omega^>(t^\prime) \, dt^\prime \nonumber \\ & &
+ \ \int^{t_0}_{-\infty}  {\cal K}^<_N(t,t^\prime) f_\omega^<(t^\prime) \,
dt^\prime.  
\label{bfkl>} \end{eqnarray}
We note that these equations are both inhomogeneous, reflecting the fact that
even for $t < t_0$, where the coupling is fixed, the BFKL equation involves 
diffusion into the running coupling regime and vice versa.

We can convert these two equations into homogeneous integro-differential
equations by operating on both sides of eq.(\ref{bfkl<}) with the operator
$${\cal O}^N_<(t) \ = \ \frac{d}{dt} \left( \frac{d}{dt}-1 \right)
   \cdots \left(\frac{d}{dt}-N \right) e^{-t/2} $$
and  operating on both sides of eq.(\ref{bfkl>}) with the 
operator
$${\cal O}^N_>(t) \ = \ \frac{d}{dt} \left( \frac{d}{dt}+1 \right)
   \cdots \left(\frac{d}{dt}+N \right) e^{t/2}. $$
These equations are $(N+1)^{th}$ order in derivatives and so each one
of these has solutions which contain $N+1$ arbitrary constants of 
integration. However, these constants are fixed by the requirement that
the original integral equations (\ref{bfkl<},\ref{bfkl>})
must be obeyed, together with the requirements
the eigenfunctions should be normalizable, which means that $f_\omega^<(t)$
must be square integrable as $t \to -\infty$ and 
 $f_\omega^>(t)$
must be square integrable as $t \to \infty$.

The general solution to the integro-differential equation for
 $f_\omega^<(t)$ is
 \beq f_\omega^<(t) \ = \  \sum_{j=1}^{N+2} a_j e^{\gamma^0_j \, t}, 
\label{mistress} \eeq
where $\gamma^0_j$ are the solutions to 
\beq b \, \omega \, t_0 \ = \ \chi_N^<(\gamma) + \chi_N^>(\gamma) \ \equiv \ 
\chi_N(\gamma)  \label{fixed} \eeq
with $\Re e (\gamma) \geq 0$ to ensure
square integrability as $t \to -\infty$.
In the allowed range of eigenvalues, eq.(\ref{range}),  there will be 
$N+2$ such solutions. For example, in the case $N=0$ the solutions are
\beq \gamma^0_j \ = \ \pm \, i \, \sqrt{\left(\lambda_0-\frac{1}{4}\right)}
\label{n0} \eeq
where 
$$\lambda_0= \frac{1}{b \, \omega \, t_0 + 2}.$$
For the allowed range, $\lambda_0 \, > \, 1/4$, and so these values are purely
imaginary. 
In the case $N=1$ we have
\beq \gamma^0_j \ = \  \pm \sqrt{ \left(
 \frac{10}{8}-2\lambda_1 \pm \sqrt{1-2 \lambda_1+4 \lambda_1^2} \right)}
\label{n1} \eeq
where 
$$\lambda_1= \frac{1}{b \, \omega \, t_0 + 3}.$$
For the allowed range, $\lambda_1 \, > \, 3/16$ and we see that two of the 
solutions are purely imaginary, $\pm i \nu_1$, and two purely real
 $\pm \beta_1$. Thus the general solution is
\beq f_\omega^<(t) \ = \ a_1 e^{i\nu_1 t} \, + \, a_2  e^{-i\nu_1 t} 
 \, + \, a_3 e^{\beta_1t}. \label{soln<1} \eeq

For the integro-differential equation for $f_\omega^>(t)$, it is convenient
to work in terms of the Laplace transform $\tilde{f}_\omega(\gamma)$ 
defined by
\beq \frac{f^>_\omega(t) }{t} \ = \ \frac{1}{2\pi i}\int d\gamma \,
\tilde{f}_\omega(\gamma) e^{\gamma t}, \label{laplace} \eeq
in which the integral over $\gamma$ is performed over some suitable contour,
 described below. The function $\tilde{f}_\omega(\gamma)$ obeys the 
differential equation
\beq \omega \, b \, \tilde{f}^{\prime\prime}_\omega(\gamma) \, + \, \chi_N(\gamma)
 \tilde{f}^\prime_\omega(\gamma)+ \chi_N^{> \, \prime} \tilde{f}_\omega(\gamma)
 \ = \ 0, \eeq 
where the prime indicates differentiation with respect to $\gamma$. 
This equation can be solved in WKB approximation yielding the result
\beq \tilde{f}_\omega(\gamma) \ = \ 
  \frac{1}{\left( V(\gamma,\omega) \right)^{1/4}} \,
\exp \left\{ -\frac{1}{2\omega b} \int^\gamma \left( \chi_N(\gamma^\prime)
+\sqrt{V(\gamma^\prime,\omega)}\right) d\gamma^\prime \right\},
\label{soln>}  \eeq
where 
$$ V(\gamma,\omega) \ = \ \chi_N^2(\gamma)-4 \, \omega \, b \, 
\chi_N^{> \, \prime}(\gamma). $$
Note that if we set the term $\chi_N^{> \, \prime}(\gamma)$ to zero we
recover the expression obtained in \cite{thorne} for the case of 
deep-inelastic scattering in which the transverse momenta are ordered and
the coupling is chosen to run always with $t$.

The solution (\ref{soln>}) is a valid approximation except near the 
points where $V(\gamma)$ vanishes. We could improve the solution by expanding
 $V(\gamma)$ to linear order around the zeroes and matching the solutions
 either side to an Airy function. However, the singularities in 
$\tilde{f}_\omega$ at these zeroes are integrable so that such an
improvement would have negligible effect on the inverse Laplace transform.

For $t<t_c$ the solution is oscillatory.
The value of this critical value $t_c$ can be estimated by approximating
the integral of eq.(\ref{laplace}) by the saddle point at
$$ 2 \, \omega \, b \, t \  = \ \chi_N(\gamma)+\sqrt{V(\gamma,\omega)}. $$ 
$t_c$ is the minimum value of $t$ for which this has a solution for
purely real $\gamma$. Again, if we were to neglect
  $\chi_N^{> \, \prime}(\gamma)$, this would occur at $\gamma=0$ and 
we have approximately
$$ t_c \ = \ \frac{1}{b \,\omega} \chi_N(0). $$  

This oscillatory part is ``matched'' to the fixed-coupling solution
by requiring consistency with the original integral equations
(\ref{bfkl<}, \ref{bfkl>}). To see how this is possible, we need to examine
in more detail the possible contours of integration for eq.(\ref{laplace}).

\noindent {\bf The $\gamma$-contour}

From the positions of the simple poles in $\chi_N^<(\gamma)$
and  $\chi_N^>(\gamma)$, we note that $\tilde{f}_\omega(\gamma)$
has branch points on the real axis at $\gamma=r+1/2$ and $\gamma=-1/2-r$ for $r=0, \cdots N$.
The branch cuts can be ``combed'' in the direction of the positive
or negative real axis in such a way as to leave a portion of the real axis
between $n-1/2$ and $n+1/2$ for which $\tilde{f}_\omega(\gamma)$ is analytic.
$n$ runs from $-N$ to $N$. 

\begin{figure}
\begin{picture}(420,450)
\put(170,420){\framebox(20,20){$\gamma$}}
\put(0,450){\line(1,0){200}} \multiput(20,450)(20,0){9}{\line(0,-1){5}}
\put(0,250){\line(1,0){200}}  \multiput(20,250)(20,0){9}{\line(0,1){5}}
\put(0,350){\line(1,0){200}}
\put(0,450){\line(0,-1){200}}   \multiput(0,440)(0,-20){10}{\line(1,0){5}}
\put(100,450){\line(0,-1){200}} \multiput(200,440)(0,-20){10}{\line(-1,0){5}}
\put(200,450){\line(0,-1){200}}
\multiput(80,253)(0,10){20}{\line(0,1){5}} \put(60,270){${\cal C}_{-1}$}
\put(0,348){\rule{70pt}{4pt}} \put(90,348){\rule{20pt}{4pt}}
 \put(110,347){\rule{20pt}{6pt}}  \put(130,346){\rule{70pt}{8pt}}
\put(15,240){-4}\put(35,240){-3}\put(55,240){-2}\put(75,240){-1}
\put(97,240){0}
\put(117,240){1}\put(137,240){2}\put(157,240){3}\put(177,240){4}

\put(390,420){\framebox(20,20){$\gamma$}}
\put(220,450){\line(1,0){200}} \multiput(240,450)(20,0){9}{\line(0,-1){5}}
\put(220,250){\line(1,0){200}}  \multiput(240,250)(20,0){9}{\line(0,1){5}}
\put(220,350){\line(1,0){200}}
\put(220,450){\line(0,-1){200}}   \multiput(220,440)(0,-20){10}{\line(1,0){5}}
\put(320,450){\line(0,-1){200}} \multiput(420,440)(0,-20){10}{\line(-1,0){5}}
\put(420,450){\line(0,-1){200}}
\multiput(317,253)(0,10){20}{\line(0,1){5}} \put(323,270){${\cal C}_{0}$}
\put(220,347){\rule{70pt}{6pt}} \put(290,348){\rule{20pt}{4pt}}
 \put(330,348){\rule{20pt}{4pt}}  \put(350,347){\rule{70pt}{6pt}}
\put(235,240){-4}\put(255,240){-3}\put(275,240){-2}\put(295,240){-1}
\put(337,240){1}\put(357,240){2}\put(377,240){3}\put(397,240){4}
\put(317,240){0}

\put(170,190){\framebox(20,20){$\gamma$}}
\put(0,220){\line(1,0){200}} \multiput(20,220)(20,0){9}{\line(0,-1){5}}
\put(0,20){\line(1,0){200}}  \multiput(20,20)(20,0){9}{\line(0,1){5}}
\put(0,120){\line(1,0){200}}
\put(0,220){\line(0,-1){200}}   \multiput(0,440)(0,-20){10}{\line(1,0){5}}
\put(100,220){\line(0,-1){200}} \multiput(200,440)(0,-20){10}{\line(-1,0){5}}
\put(200,220){\line(0,-1){200}}
\multiput(120,23)(0,10){20}{\line(0,1){5}} \put(107,40){${\cal C}_{1}$}
\put(70,117){\rule{20pt}{6pt}} \put(90,118){\rule{20pt}{4pt}}
 \put(0,116){\rule{70pt}{8pt}}  \put(130,116){\rule{70pt}{8pt}}
\put(15,10){-4}\put(35,10){-3}\put(55,10){-2}\put(75,10){-1}
\put(97,10){0}
\put(117,10){1}\put(137,10){2}\put(157,10){3}\put(177,10){4}

\put(390,190){\framebox(20,20){$\gamma$}}
\put(220,220){\line(1,0){200}} \multiput(240,220)(20,0){9}{\line(0,-1){5}}
\put(220,20){\line(1,0){200}}  \multiput(240,20)(20,0){9}{\line(0,1){5}}
\put(220,120){\line(1,0){200}}
\put(220,220){\line(0,-1){200}}   \multiput(220,440)(0,-20){10}{\line(1,0){5}}
\put(320,220){\line(0,-1){200}} \multiput(420,440)(0,-20){10}{\line(-1,0){5}}
\put(420,220){\line(0,-1){200}}
\multiput(360,23)(0,10){20}{\line(0,1){5}} \put(344,40){${\cal C}_{2}$}
\put(220,116){\rule{90pt}{8pt}}  \put(220,115){\rule{70pt}{10pt}}
 \put(330,118){\rule{20pt}{4pt}}  \put(310,117){\rule{20pt}{6pt}}

\put(235,10){-4}\put(255,10){-3}\put(275,10){-2}\put(295,10){-1}
\put(337,10){1}\put(357,10){2}\put(377,10){3}\put(397,10){4}
\put(317,10){0}

\end{picture}
\caption{The four contours for the case $N=1$ corresponding to the four
different ways of ``combing'' the branch cuts.}
\label{fig1} \end{figure}
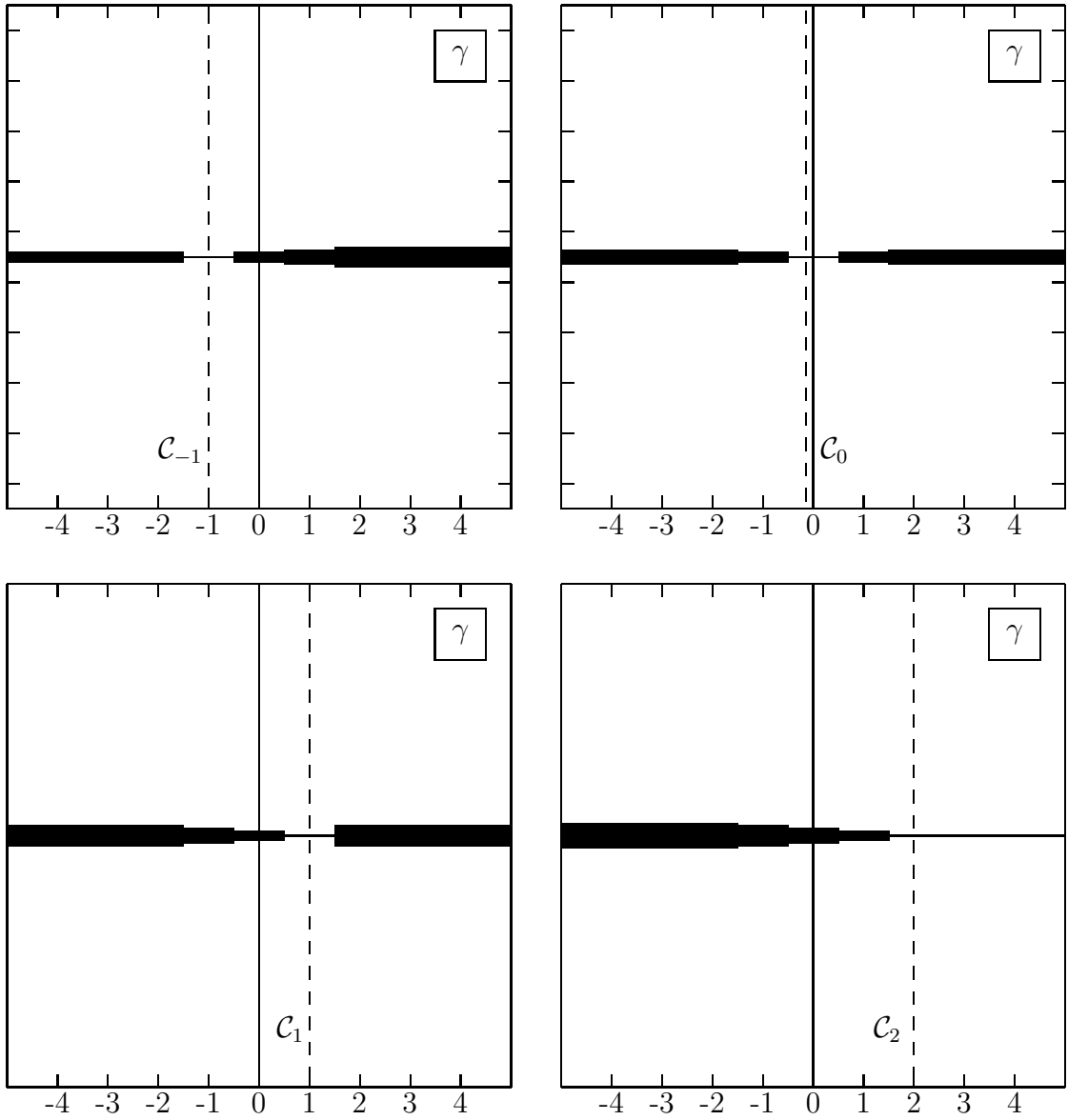

For each such ``combing'' a contour, ${\cal C}_n$
can be chosen that crosses 
the real axis between $n-1/2$ and
 $n+1/2$. Since $t>0$ we require that the ends of the contour
turn over so that at the ends of the contour $\Re e \,  \gamma \to -\infty$,
thus ensuring that the inverse Laplace transform integral exists. 
The integral over each such contour is a valid solution to the 
integro-differential equation for $f^>_\omega(t)$ and since this equation 
is linear, the most general solution which is square integrable as  
$t \to \infty$ is the sum of the integrals over these contours with 
$-N \, \le \, n \, \le 0$, with arbitrary coefficients $b_n$, i.e.
\beq \sum_{n=-N}^0 b_n \int_{{\cal C}_n} d \gamma \tilde{f}_\omega(\gamma)
 e^{\gamma t}, \label{master} \eeq
with $\tilde{f}_\omega(\gamma)$ given by eq.(\ref{soln>}).\footnote{We 
have absorbed the factor of $2\pi i$ in the denominator
in eq.(\ref{laplace}) into the definition of the coefficients $b_n$.}
As an example we consider the $N=1$ case. The four ``combings'' are shown
in Fig. \ref{fig1}  together with the four contours,  ${\cal C}_n$.
Only  
 ${\cal C}_{-1}$  and  ${\cal C}_0$ give a square integrable solution,
so the general solution is
\beq
f^>_\omega(t) \ = t \left\{ \ b_{-1} 
\, \int_{-\infty}^\infty d \nu \, e^{-t} \, \tilde{f}_\omega(i\nu-1)
e^{i\nu t} \ +
 \  b_0 \,\int_{-\infty}^\infty d \nu  \tilde{f}_\omega(i\nu)
e^{i\nu t}
\right\}\label{soln>1}. \eeq 

Returning now to the original equations (\ref{bfkl<}) and (\ref{bfkl>}),
we see that these are obeyed by the solutions
(\ref{master}) and (\ref{mistress}) provided
\beq
\sum_{j=1}^{N+2} a_j \frac{e^{\gamma^0_j t_0}}{(\gamma^0_j-r-1/2)}
-t_0 \sum_{n=-N}^0 b_n \int_{{\cal C}_n} d \gamma 
\frac{ \tilde{f}_\omega(\gamma) \, e^{\gamma t_0}}{(\gamma-r-1/2)}
 \ = \ 0 \ \ \ (r=0 \cdots N),
\label{match1} \eeq
\beq
\sum_{j=1}^{N+2} a_j \frac{e^{\gamma^0_j t_0}}{(\gamma^0_j+r+1/2)}
+ \sum_{n=-N}^0 b_n \int_{{\cal C}_n} d \gamma 
\frac{ \tilde{f}_\omega^\prime(\gamma) \, e^{\gamma t_0}}{(\gamma+r+1/2)}
 \ = \ 0 \ \ \ (r=0 \cdots N).
\label{match2} \eeq
Thus we obtain $2N+2$ relations between the $N+2$ coefficients $a_j$
and the $N+1$ coefficients $b_k$ (the overall normalization of these
coefficients is determined by the normalization condition of the
eigenfunctions). This then determines uniquely the required
eigenfunctions.

Equations (\ref{bfkl<}, \ref{bfkl>}) guarantee the continuity of the
eigenfunctions at $t=t_0$, i.e. $f^<_\omega(t_0)=f^>_\omega(t_0)$,
but the derivative is not continuous. 
This results from the fact that we have taken an expression for the coupling 
as a function of $t$, whose derivative is not continuous at $t=t_0$.
\medskip

We can now see how the inverse Mellin transform, $f(s,t)$ of the solution
$f_\omega(t)$ extrapolates smoothly between the power-like behaviour
for $t \sim t_0$ and the softer  behaviour (with at most logarithmic
dependence on $s$) for  $t \gg t_0$.

Examining the $\gamma-$ integral for one particular contour, 
${\cal C}_n$, and for 
$t > t_0$,
we see that this will be dominated by a saddle point
 $\gamma_s(\omega,t)$, which
is the solution to  
\beq 2 b \, \omega \, t \ = \ \chi_N(\gamma)+\sqrt{V(\gamma,\omega)}. 
\label{saddle} \eeq
This is  multi-valued and the dominant saddle point will be the one closest
to the integration contour. We call this saddle point $\gamma^n_s(\omega,t)$.
In other words, $f^>_\omega(t)$ of eq.(\ref{laplace})
from the contour ${\cal C}_n$ will contain an
$\omega-$dependent prefactor, which may be written
 (after a suitable integration by parts)
$$ \exp\left\{ \int^t
 \gamma^n_s(\omega,t^\prime) dt^\prime \right\}. $$
The essential singularity at $\omega=0$ is now encoded
 in  $\gamma^n_s(\omega,t)$  which 
acquires a singularity as $\omega \to 0$,  i.e. the R.H.S. of (\ref{saddle})
can only tend to  zero as $|\gamma| \to \infty$.

Now examining  eqs.(\ref{match1}, \ref{match2}), we see that the
contour integrals multiplying the coefficients $b_n$ are also dominated
by a similar saddle point but with $t$ set to $t_0$. Thus these coefficients
appear in  eqs.(\ref{match1}, \ref{match2}) with a prefactor
$$ \exp\left\{ \int^{t_0}
 \gamma^n_s(\omega,t^\prime) dt^\prime \right\}. $$
Therefore these coefficients also possess an essential singularity
at $\omega=0$.
Furthermore $b_n$ will contain factors of the form
 $\exp(\gamma^j_0(\omega) t_0)$
 from the L.H.S. of  (\ref{match1}, \ref{match2}).

For $t$ close to $t_0$ each term in the sum of eq.(\ref{master}),
i.e. the product of $b_n$ and the contribution to
 $f^>_\omega(t)$ from the integral over the contour ${\cal C}_n$
will have a dominant $\omega-$dependence  of the form
\beq \exp\left\{ \gamma^n_s(\omega,t_0)(t-t_0)+\gamma_0^j(\omega)t_0 \right\}.
\label{neweq} \eeq
 Note that due to the partial cancellation, the coefficient of the
term that gives rise to the essential singularity vanishes
as $t \to t_0$.

For sufficiently large $s$, the inverse Mellin transform probes
small $\omega$, where the term
in the exponent  proportional to $\gamma^n_s(\omega,t_0)$ will dominate
even if $(t-t_0)$ is small and the $s$-behaviour is dominated
by the essential singularity in the Laplace transform,
 $\tilde{f}_\omega(\gamma)$.
However, for more moderate values of $s$  we can neglect this term
when $t$ is sufficiently close to $t_0$ and the
$s-$dependence will be dominated by the fixed coupling power-like behaviour.
As $t$ is increased away from $t_0$, the boundary in $s$ where the
soft behaviour takes over from the power-like behaviour decreases and we see
 that as $t \to \infty$ we recover the solution of \cite{thorne}. 

\begin{figure} 
\centerline{\epsfig{file=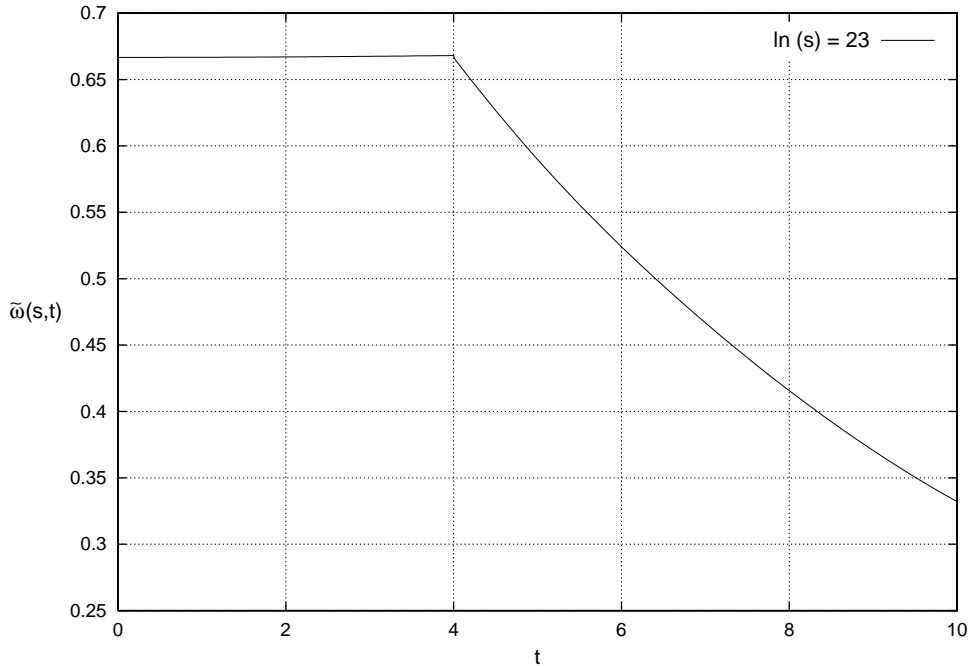,width=9cm,angle=270}}
\caption{$\tilde{\omega}(s,t)$ against $t$ for  $\ln s=23$}
\label{fig2}
\end{figure}

This formalism is considerably simplified if we assume that $t_0$
is sufficiently large that 
 $\exp(-t_0) \ll 1$. In this case we only need
consider the right-most contour consistent with square-integrability
as $t \to \infty$. This is the contour ${\cal C}_0$. In this case,
within the approximation of performing the $\gamma-$contour integral
and the inverse Mellin transform by the saddle-point method, the
$s-$dependence may be written as $s^{\tilde{\omega}(s,t)},$
where  $\tilde{\omega}(s,t)$ is the solution to 
\beq 
\ln(s) \ = \  - \, \frac{d}{d\omega} \left\{
 \theta(t-t_0) \int_{t_0}^t \gamma^0_s(\omega,t^\prime) dt^\prime
 + \gamma_0^m(\omega) {\rm Min}(t,t_0) \right\} \label{plot}, \eeq 
and $\gamma_0^m(\omega)$ is the solution to eq.(\ref{fixed})
with the largest real part.
In Fig. \ref{fig2} we plot 
$\tilde{\omega}(s,t)$ against $t$ for a large  value of $s \ (\ln s =23)$. 
For simplicity we have taken the $N=0$ case, although we expect
the result to be qualitatively the same for higher values of $N$,
and we take $t_0=4$.
We see that it is almost constant for
 $t \, < \, t_0$ and diminishes  as one increases $t$ from $t_0$,
indicating the smooth transition from hard to soft $s-$dependence
as $t$ is increased. The discontinuity of the slope at $t=t_0$
is a reflection of the fact that we have taken a sharp discontinuity
in the slope of  the running  coupling at $t_0$ and would be smoothed out
if a smoother transition were taken.

\begin{figure} 
\centerline{\epsfig{file=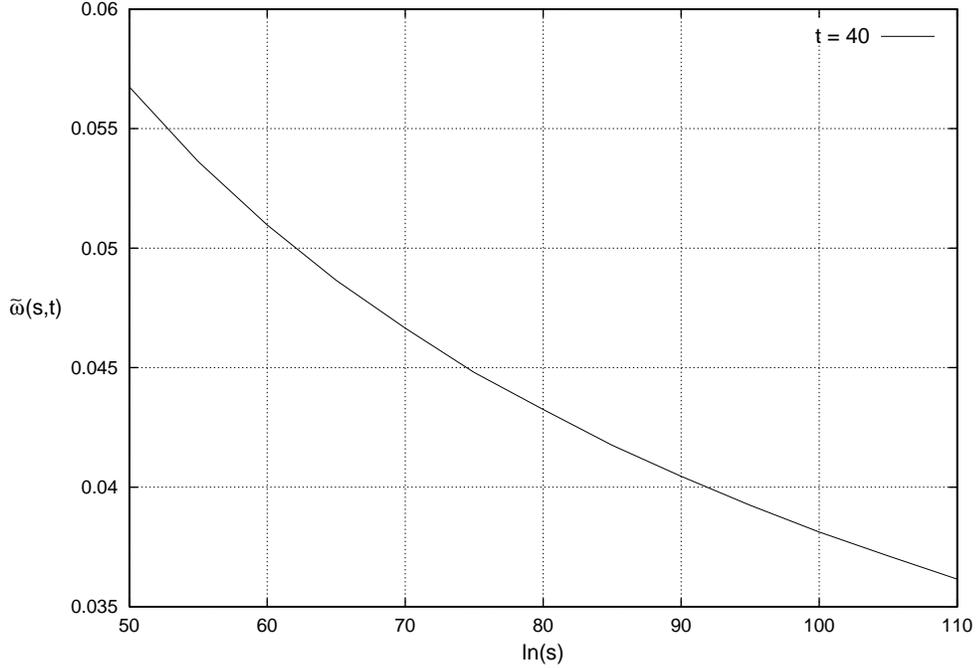,width=9cm,angle=270}}
\caption{$\tilde{\omega}(s,t)$ against $\ln s$ for $t=40$}
\label{fig3}
\end{figure}

The asymptotic behaviour is only obtained for much larger values
of $t$ and in Fig. \ref{fig3} we show the $s$ dependence 
of $\tilde{\omega}(s,t)$ for $t=40$
and notice that it is much closer to zero than the case
where $t \sim t_0$ and  decreases as $\ln s$
increases, resulting in an $s-$dependence for the amplitude
that has no fixed power of $s$.  

\noindent {\bf Accounting for Higher Order Corrections} 

The higher order BFKL kernel \cite{BFKL2} does not lend itself readily
to an expansion in  powers of $e^{t-t^\prime}$. Furthermore, it is known
that these corrections are large. 

However, it has been pointed out \cite{salam} 
that a substantial part of the higher order corrections 
consist of ``collinear corrections'' which are 
required to ensure correct behaviour when the formalism is applied
to deep inelastic scattering. It was shown in \cite{salam} and confirmed
in \cite{us} that once these corrections are accounted for, the remaining
higher order corrections are modest. 
The simplest way of encoding these collinear corrections and summing them
is to replace $\gamma$  by $(\gamma-\omega/2)$ in $\chi_N^<(\gamma)$
and by $(\gamma+\omega/2)$ in $\chi_N^>(\gamma)$ \cite{ags,salam}.

This formalism is readily adapted to the procedure described here.
In the expression (\ref{mistress}) for   $f_\omega^<(t)$
the quantities  $\gamma_j^0(\omega)$ are taken to be the solutions
 to the
implicit equation
\beq  \chi_N^<(\gamma-\omega/2)+\chi_N^>(\gamma+\omega/2)
\ = \ \omega b t_0, \label{gamm1} \eeq
and the expression for the Laplace transform of   $f_\omega^>(t)$
becomes
\beq  
\tilde{f}_\omega(\gamma) \ = \ \frac{1}{(V(\gamma,\omega))^{1/4}}
 \, \exp\left\{-\frac{1}{2\omega b} \int^\gamma \left(
 \chi_N^<(\gamma^\prime-\omega/2)+\chi_N^>(\gamma^\prime+\omega/2)
+\sqrt{V(\gamma^\prime,\omega)} \right) d\gamma^\prime
 \right\}, \label{soln2>}\eeq
where 
\beq V(\gamma,\omega) \ = \ \left(\chi_N^<(\gamma-\omega/2)
+\chi_N^>(\gamma+\omega/2)\right)^2
-4 \omega b \,  
 \chi_N^{> \, \prime} (\gamma+\omega/2). \eeq  
Provided that we restrict our analysis to the case $|\omega|<1$, the
singularity structure and consequently the analysis of the contours
of integration remains unchanged from the leading order case,  
although the exact positions of the branch points will 
have moved.

\bigskip

\noindent {\bf Acknowledgement}
ASV wishes to thank the University of Manchester and the CERN Theory
Division where this work started and acknowledges the support of PPARC 
(Posdoctoral Fellowship: PPA/P/S/1999/00446). DAR wishes to thank
the CERN Theory Division for its hospitality.

\end{document}